\documentclass[12pt]{article}
\usepackage{amsmath,amsfonts,amssymb}

\begin{document}

\begin{titlepage}

\begin{center}

\vskip 0.4 cm

\begin{center}
{\Large  \bf Non-Projectable Ho\v{r}ava-Lifshitz Gravity without
Unwanted Scalar Graviton }
\end{center}

\vskip 1cm

\vspace{1em} Masud Chaichian,$^a$ Josef Kluso\v{n},$^b$ and Markku
Oksanen,$^{a,}$ \footnote{Email addresses:
masud.chaichian@helsinki.fi (M. Chaichian), klu@physics.muni.cz (J.
Kluso\v{n}), markku.oksanen@helsinki.fi (M. Oksanen)
}\\
\vspace{1em}$^a$\textit{Department of Physics, University of Helsinki,
P.O. Box 64,\\ FI-00014 Helsinki, Finland}\\
\vspace{.3em} $^b$\textit{Department of Theoretical Physics and
Astrophysics, Faculty of Science,\\
Masaryk University, Kotl\'a\v{r}sk\'a 2, 611 37, Brno, Czech Republic}

\vskip 0.8cm

\end{center}

\begin{abstract}
We consider a way of eliminating the unwanted scalar graviton from
Ho\v{r}ava-Lifshitz gravity. That is achieved via introduction of
certain additional constraints. We perform canonical analysis of both
projectable and non-projectable versions of the theory. We obtain the
structure of constraints in each case, and analyze its dependence on
the values of the coupling constants involved in the additional
constraints. In the non-projectable theory, the scalar graviton is
absent when the coupling constants have certain values,
while for other values the scalar graviton appears.
The projectable theory is free from the scalar graviton regardless of
the values of the coupling constants, even though the structure of
constraints does depend on the value of a coupling constant.
\end{abstract}

\end{titlepage}
\bigskip

\newpage

\def\bn{\mathbf{n}}
\newcommand{\bC}{\mathbf{C}}
\newcommand{\bD}{\mathbf{D}}
\def\hf{\hat{f}}
\def\tK{\tilde{K}}
\def\tmG{\tilde{\mG}}
\def\mC{\mathcal{C}}
\def\bk{\mathbf{k}}
\def\tp{\tilde{p}}
\def\tr{\mathrm{tr}\, }
\def\tmH{\tilde{\mH}}
\def\tY{\mathcal{Y}}
\def\nn{\nonumber \\}
\def\bI{\mathbf{I}}
\def\tPhi{\tilde{\Phi}}
\def\tmV{\tilde{\mV}}
\def\e{\mathrm{e}}
\def\bE{\mathbf{E}}
\def\bX{\mathbf{X}}
\def\bY{\mathbf{Y}}
\def\bR{\bar{R}}
\def\hN{\hat{N}}
\def\hK{\hat{K}}
\def\hnabla{\hat{\nabla}}
\def\hc{\hat{c}}
\def\mH{\mathcal{H}}
\def \Gi{\left(G^{-1}\right)}
\def\hZ{\hat{Z}}
\def\tpi{\tilde{\pi}}
\def\bz{\mathbf{z}}
\def\bK{\mathbf{K}}
\def\iD{\left(D^{-1}\right)}
\def\tmJ{\tilde{\mathcal{J}}}
\def\tr{\mathrm{Tr}}
\def\mJ{\mathcal{J}}
\def\partt{\partial_t}
\def\parts{\partial_\sigma}
\def\bG{\mathbf{G}}
\def\str{\mathrm{Str}}
\def\Pf{\mathrm{Pf}}
\def\bM{\mathbf{M}}
\def\tA{\tilde{A}}
\newcommand{\mW}{\mathcal{W}}
\def\bx{\mathbf{x}}
\def\by{\mathbf{y}}
\def \mD{\mathcal{D}}
\newcommand{\tZ}{\tilde{Z}}
\newcommand{\tW}{\tilde{W}}
\newcommand{\tmD}{\tilde{\mathcal{D}}}
\newcommand{\tN}{\tilde{N}}
\newcommand{\hC}{\hat{C}}
\newcommand{\hg}{g}
\newcommand{\hX}{\hat{X}}
\newcommand{\bQ}{\mathbf{Q}}
\newcommand{\hd}{\hat{d}}
\newcommand{\tX}{\tilde{X}}
\newcommand{\calg}{\mathcal{G}}
\newcommand{\calgi}{\left(\calg^{-1}\right)}
\newcommand{\hsigma}{\hat{\sigma}}
\newcommand{\hx}{\hat{x}}
\newcommand{\tchi}{\tilde{\chi}}
\newcommand{\mA}{\mathcal{A}}
\newcommand{\ha}{\hat{a}}
\newcommand{\tB}{\tilde{B}}
\newcommand{\hrho}{\hat{\rho}}
\newcommand{\hh}{\hat{h}}
\newcommand{\homega}{\hat{\omega}}
\newcommand{\mK}{\mathcal{K}}
\newcommand{\hmK}{\hat{\mK}}
\newcommand{\hA}{\hat{A}}
\newcommand{\mF}{\mathcal{F}}
\newcommand{\hmF}{\hat{\mF}}
\newcommand{\hQ}{\hat{Q}}
\newcommand{\mU}{\mathcal{U}}
\newcommand{\hPhi}{\hat{\Phi}}
\newcommand{\hPi}{\hat{\Pi}}
\newcommand{\hD}{\hat{D}}
\def\tmC{\tilde{\mC}}
\newcommand{\hb}{\hat{b}}
\def\I{\mathbf{I}}
\def\tW{\tilde{W}}
\newcommand{\tD}{\tilde{D}}
\newcommand{\mG}{\mathcal{G}}
\def\IT{\I_{\Phi,\Phi',T}}
\def \cit{\IT^{\dag}}
\newcommand{\hk}{\hat{k}}
\def \cdt{\overline{\tilde{D}T}}
\def \dt{\tilde{D}T}
\def\bra #1{\left<#1\right|}
\def\ket #1{\left|#1\right>}
\def\mV{\mathcal{V}}
\def\Xn #1{X^{(#1)}}
\newcommand{\Xni}[2] {X^{(#1)#2}}
\newcommand{\bAn}[1] {\mathbf{A}^{(#1)}}
\def \bAi{\left(\mathbf{A}^{-1}\right)}
\newcommand{\bAni}[1]
{\left(\mathbf{A}_{(#1)}^{-1}\right)}
\def \bA{\mathbf{A}}
\newcommand{\bT}{\mathbf{T}}
\def\bmR{\bar{\mR}}
\newcommand{\mL}{\mathcal{L}}
\newcommand{\mbQ}{\mathbf{Q}}
\def\mat{\tilde{\mathbf{a}}}
\def\mtF{\tilde{\mathcal{F}}}
\def \tZ{\tilde{Z}}
\def\mtC{\tilde{C}}
\def \tY{\tilde{Y}}
\def\pb #1{\left\{#1\right\}}
\newcommand{\E}[3]{E_{(#1)#2}^{ \quad #3}}
\newcommand{\p}[1]{p_{(#1)}}
\newcommand{\hEn}[3]{\hat{E}_{(#1)#2}^{ \quad #3}}
\def\mbPhi{\mathbf{\Phi}}
\def\tg{\tilde{g}}
\newcommand{\phys}{\mathrm{phys}}

\section{Introduction}
In 2009 P. Ho\v{r}ava introduced power-counting renormalizable
theory of gravity  in \cite{Horava:2009uw} \footnote{For review, see
\cite{Weinfurtner:2010hz,Sotiriou:2010wn}.}. This theory is
renormalizable thanks to the existence of the anisotropic scaling
\begin{equation}
t\rightarrow b^z t \ , \bx\rightarrow b \bx \ ,
\end{equation}
where $z$ is dynamical critical exponent where $z\geq 3$ in the
space-time with three spatial dimensions. This theory is now known
as Ho\v{r}ava-Lifshitz (HL) gravity.  The physical variables in HL
gravity follow from standard $3+1$ decomposition of the metric
\cite{Arnowitt:1962hi} \footnote{For review, see
\cite{Gourgoulhon:2007ue}.} and consists from the lapse $N$, shift
$N^i$ and the three-dimensional spatial metric $g_{ij}$. Currently
there are two version of HL gravity. In the first one, known as
projectable theory, the lapse depends on time only:
\begin{equation}
N=N(t) \ , \quad N^i=N^i(t,\bx) \ , \quad g_{ij}=g_{ij}(t,\bx) \ .
\end{equation}
On the other hand in non-projectable theory the lapse depends on the
spatial coordinates as well
\begin{equation}
N=N(t,\bx) \ , \quad N^i=N^i(t,\bx) \ , \quad g_{ij}=g_{ij}(t,\bx) \
.
\end{equation}
Since space and time coordinates scale differently in HL gravity,
the theory is not invariant under full four-dimensional diffeomorphism
but only under so-called foliation preserving diffeomorphism
\begin{equation}
t\rightarrow t'(t) \ , \quad \bx\rightarrow \bx'(t,\bx) \ .
\end{equation}
Due to the fact that the number of gauge symmetries is restricted
the number of propagating degrees of freedom is larger. Explicitly,
the theory contains non-only a tensor graviton but also a scalar
graviton and the consistency of the theory crucially depends on the
properties of the scalar graviton. The properties of given scalar
graviton depends on the fact whether we consider projectable or
non-projectable HL gravity where it turns out that projectable
theory suffers from infrared instability. It turns out that the
scalar graviton has much better properties when we consider
non-projectable theory  \cite{Blas:2009qj,Blas:2010hb}.

Another possibility how to avoid the problems with the scalar graviton
was proposed by P. Ho\v{r}ava and Melby-Thompson in \cite{Horava:2010zj}
where $U(1)$ extension of the projectable version of HL gravity was
considered. Thanks to this additional symmetry we can argue that the
scalar graviton is absent. It was argued originally that the
presence of given symmetry fixed the dimensionless parameter
$\lambda$ that appears in the definition of the generalized de Witt
metric to be equal to one. However, it was shown in
\cite{daSilva:2010bm} that this $U(1)$ symmetry is preserved for any
value of $\lambda$. The absence of the scalar graviton when
$\lambda\neq1$, and the potential problems regarding stability, ghosts
and strong coupling, have been analyzed in \cite{Huang:2010ay}.
The Hamiltonian analysis of projectable HL gravity with the extra
$U(1)$ symmetry
was also performed in \cite{Kluson:2010zn}. Moreover, it was argued in
\cite{Kluson:2011xx} that the same number of degrees of freedom can
be found in the Lagrange multiplier modified HL gravity which
implies an existence of the additional constraint. However, the
condition of the preservation of given constraint during the time
development of the system implies an additional constraint which is
more complicated and it is very difficult to solve it explicitly.
Further, the symplectic structure of the Lagrange multiplier
modified HL gravity is also very involved.
Then one can ask the question whether it is possible to
formulate HL gravity with additional constraints that can be
explicitly solved and with simpler symplectic structure. The aim of
this paper is to construct such a form of the non-projectable HL
gravity with two additional constraints. The Hamiltonian analysis of
non-projectable HL gravity was performed in \cite{Kluson:2010nf},
see also \cite{Bellorin:2010te,Chaichian:2010zn}. This analysis was
further extended in \cite{Donnelly:2011df} with very important
identifications of the global first class constraints whose analysis
was missing in \cite{Kluson:2010nf}. Recently the Hamiltonian analysis
of non-projectable HL gravity with $U(1)$ symmetry was studied  in
\cite{Mukohyama:2015gia} where the structure of local constraints was
very carefully analyzed.

 In this paper we consider
more general situation when we have two additional constraints in
the non-projectable and projectable gravity. Following
\cite{Mukohyama:2015gia} we consider the most general form of these
additional constrains that preserve the power counting
renormalizability of HL gravity. Then we argue that for the generic
form of the parameters that define these constraints the additional
constrains cannot eliminate the scalar graviton.
 On the other hand
we find that in some exceptional cases this scalar graviton can be
eliminated. This analysis can be considered as the generalization of
the analysis presented in \cite{Mukohyama:2015gia}. We also discuss
the form of two global first class constraints which were not
analyzed in \cite{Mukohyama:2015gia}. These constraints cannot
eliminate local degrees of freedom but reflect the invariance of the
theory under foliation preserving diffeomorphism
\cite{Donnelly:2011df}.

It has been observed that the linearized approximation of HL gravity
with a truncated potential (that consists of only the
nonrenormalizable terms ${}^{(3)}\!R$, ${}^{(3)}\!R^2$ and
${}^{(3)}\!R_{ij}{}^{(3)}\!R^{ij}$) does not contain the scalar
graviton \cite{Das:2011tx}. However, the extra scalar mode is known to
be present in the corresponding nonlinear theory
\cite{Bellorin:2010te}. In order to avoid missing any physical degrees
of freedom, we shall perform our analysis within a full nonlinear
theory.

The structure of given paper is as follows. In section~\ref{second}
we introduce the non-projectable HL gravity with two additional
constraints. Then in section~\ref{third} we perform Hamiltonian analysis
of the introduced theory with dependence on the values of the
parameters that appear in these constraints. In section~\ref{fourth} we
perform canonical analysis of projectable HL gravity with additional
constraints. Finally in conclusion~\ref{fifth} we outline our results.

\section{Non-projectable HL gravity with additional
constraints}\label{second}
In this section we will propose  non-projectable HL gravity with
additional constraint. Explicitly, we consider the action in the
form
\begin{eqnarray}\label{Snoproj}
S&=&\frac{1}{\kappa^2} \int dt d^3\bx [N\sqrt{g}(K_{ij}
\mG^{ijkl}K_{kl}-\mV(g,a_i)+\Lambda \sqrt{g}K+
\frac{1}{4}\sqrt{g}f(\Lambda)g^{ij}\mG_{ijkl}g^{kl} \nonumber \\
&+& A\sqrt{g}({}^{(3)}R-\Omega+\eta_1 a_i a^i+\eta_2 \nabla_i a^i)] \ ,
\end{eqnarray}
where  $N=N(\bx,t), a_i=\frac{\nabla_i N}{N}$ and $K_{ij}$ is equal
to
\begin{equation}
K_{ij}=\frac{1}{2N}(\partial_t g_{ij}-\nabla_i N_j-\nabla_j N_i) \ .
\end{equation}
Further the generalized De Witt metric $\mG^{ijkl}$ is
defined as
\begin{equation}
\mG^{ijkl}=\frac{1}{2}(g^{ik}g^{jl}+ g^{il}g^{jk})-\lambda
g^{ij}g^{kl} \
\end{equation}
with inverse
\begin{equation}
\mG_{ijkl}=\frac{1}{2}(g_{ik}g_{jl}+
g_{il}g_{jk})-\frac{\lambda}{3\lambda-1} g_{ij}g_{kl}  \ , \quad
\mG_{ijkl}\mG^{klmn}= \frac{1}{2}(\delta_i^m\delta_j^n+\delta_i^n
\delta_j^m) \ .
\end{equation}
Finally note that the generic potential of non-projectable HL has
the form
\begin{eqnarray}
\mV(g_{ij},a)&=&\rho^2 \alpha_1-\alpha_2{}^{(3)}R-\alpha_3 a_i a^i
\nonumber \\
&+&\rho^{-2}(\beta_1 {}^{(3)}R^2+\beta_2
{}^{(3)}R^{ij}{}^{(3)}R_{ij}+
\beta_3 \nabla^2 {}^{(3)}R \nonumber \\
&+&\beta a_4 a_i\nabla^2 a^i+\beta_5 (\nabla_i a^i)^2+\beta_6 (a_i
a^i)^2+\beta_7 a_i a_j {}^{(3)}R^{ij}+\dots) \nonumber \\
&+&\rho^{-4}(\omega_1 {}^{(3)}R^3+\omega_2
{}^{(3)}R{}^{(3)}R^{ij}{}^{(3)}R_{ij}+\omega_3 {}^{(3)}R^i_{ \ j}
{}^{(3)}R^j_{ \ k}{}^{(3)}R^k_{ \ i} \nonumber \\
&+&\omega_4 \nabla^i{}^{(3)}R^{jk}\nabla_i {}^{(3)}R_{jk}+\omega_5
{}^{(3)}R \nabla^2 {}^{(3)}R \nonumber \\
&+&\omega_6 \nabla^4 {}^{(3)}R+\omega_7 a_i \nabla^4a^i+\omega_8 a_i
a^i a_j \nabla^2 a^j \nonumber \\
&+&\omega_9 (a_i a^i)^3+\omega_{10} a_i a^i a_j a_k {}^{(3)}R^{jk}+
\omega_{11} a_i a^j {}^{(3)}R^{ik}{}^{(3)}R_{jk}+\dots) \ ,
\end{eqnarray}
where $\rho^2=\frac{1}{2\kappa}$, where $\kappa^2$ is the constant
that reduces to the gravitational constant at low energy. Further,
$\alpha_i,\beta_i$ and $\omega_i$ are dimensionless coupling
constants.
 We also introduced the general function of Lagrange
multiplier $f(\Lambda)$ whose specific form will be determined
later.

Before we proceed to the Hamiltonian formulation of the theory
introduced above we
should explain the presence of the terms with the constants
$\eta_1,\eta_2$. To do this we
 list the scaling dimensions of coordinates
and fields (in mass units)
\begin{eqnarray}
[t]&=&-z \ , \quad [x^i]=-1 \ , \quad [dt d^3\bx]=-z-3 \ ,
\nonumber \\
\left[g_{ij}\right]&=&0 \ ,\quad [N]=0 \ ,\quad
[N_i]=[N^i]=z-1 \ , \nonumber \\
\left[K_{ij}\right]&=&z \ ,\quad \left[{}^{(3)}R\right]=2 \ , \quad
[A]=2z-2 \ , \quad [\Lambda]=z \ .
\end{eqnarray}
Observe that the scaling dimensions of the kinetic terms are
$[K^{ij}K_{ij}]=[K^2]=2z$ and hence power counting renormalizability
requires that the other terms in the action should have scaling
dimensions equal to or less than $2z$. Then we observe that the
expression $A {}^{(3)}R$ is marginal with the scaling dimension
$2z$. We also see that
\begin{equation}
[A^n]=2n(z-1)>2z \ , \quad \mathrm{for} \  n\geq 2 \ , z\geq 3
\end{equation}
and hence in order to preserve renormalizability of the action it
should contain terms at most linear in $A$. Note also that the
spatial derivatives of all other fundamental variables have positive
scaling dimensions. Further, from the fact that the scaling
dimension of $\Lambda$ is $[\Lambda]=z$ we see that $f(\Lambda)$
could be quadratic function. Since $[a_i]=1$ we see that generally
there could be terms linear in $\Lambda$ that is multiplied with
$a_ia^i$. Explicitly we can presume that $f(\Lambda)$ has the form
\begin{equation}
f(\Lambda)=\gamma_0\Lambda+\gamma_1 a_i a^i\Lambda
+\gamma_2 \nabla_i a^i\Lambda+\gamma_3\Lambda^2 \ ,
\end{equation}
where $\gamma_0,\gamma_1$ and $\gamma_2$ are dimensional
constants, and $\gamma_3$ is a dimensionless constant, as the dimension
of $f(\Lambda)$ is $2z$.
We will show that the Hamiltonian structure of the theory crucially
depends on specific values of these constants. However, from the form
of the function $f(\Lambda)$ and from the action (\ref{Snoproj}) it
is clear that the terms that are multiplied by the constants $\eta_2
$ and $\gamma_2$ have the same impact on the constraint structure as
terms multiplied by $\eta_1,\gamma_1$. For simplicity of the
resulting expressions we will presume that $\gamma_2=\eta_2=0$
keeping in mind that the same analysis is valid for the general case
as well.

\section{Hamiltonian formalism}\label{third}
Now we are ready to proceed to the Hamiltonian formulation.
 From the action (\ref{Snoproj}) we obtain
\begin{eqnarray}
\pi^{ij}&=&\frac{\delta \mL}{\delta
\partial_t g_{ij}}=\frac{1}{\kappa^2}\sqrt{g}\mG^{ijkl}K_{kl}
 +\frac{1}{2\kappa^2}\Lambda \sqrt{g}g^{ij} \ , \nonumber \\
\pi_N&=&\frac{\delta \mL}{\delta \partial_t N}=0 \ , \quad \pi_i=
\frac{\delta \mL}{\delta \partial_t N^i}=0 \ , \nonumber \\
 p_A&=&\frac{\delta \mL}{\delta \partial_t A}=0 \ , \quad
 p_\Lambda=\frac{\delta \mL}{\delta \partial_t \Lambda}=0 \ ,
\end{eqnarray}
and hence we  find the bare Hamiltonian in the form
\begin{eqnarray}
H_B&=&\int d^3\bx (\pi^{ij}\partial_t g_{ij}-\mL) \nonumber \\
&=& \int dt d^3\bx [N\frac{\kappa^2}{\sqrt{g}}
(\pi^{ij}-\frac{1}{2}\Lambda \sqrt{g}g^{ij}) \mG_{ijkl}
(\pi^{kl}-\frac{1}{2}\Lambda \sqrt{g}g^{kl}) \nonumber \\
&+&\frac{1}{\kappa^2}N\sqrt{g}\mV(g,a_i)
-\frac{1}{\kappa^2}N\sqrt{g}(\gamma_0\Lambda+\gamma_1 a_i
a^i\Lambda
+\gamma_3\Lambda^2)g^{ij}\mG_{ijkl}g^{kl} \nonumber \\
&-&\frac{1}{\kappa^2}AN\sqrt{g}({}^{(3)}R-\Omega+\eta_1 a_i a^i
 )+N^i\mH_i]
\end{eqnarray}
together with following set of primary constraints
\begin{equation}
\pi_N\approx 0 \ , \quad  \pi_i\approx 0 \ , \quad p_A\approx 0 \ ,
\quad p_\Lambda \approx 0 \ .
\end{equation}
It turns out that there is a particular combination of the primary
constraint $\pi_N$ that is defined as
\cite{Donnelly:2011df}
\begin{equation}
\Pi_N=\int d^3\bx N\pi_N \ .
\end{equation}
This constraint obeys the relation
\begin{equation}
\pb{\Pi_N,a_i(\bx)}=0
\end{equation}
and also
\begin{equation}
\pb{\Pi_N,N(\bx)}=N(\bx) \ ,\quad  \pb{\Pi_N,\pi_N(\bx)}=-\pi_N(\bx) \ .
\end{equation}
In the usual non-projectable HL gravity, $\Pi_N$ is a first class
constraint.  Hence we have to be careful with the definition of the
local and global constraints.
It is instructive to define the following local constraint
\begin{equation}
\tilde{\pi}_N(\bx)=\pi_N(\bx)-\frac{\sqrt{g}(\bx)}{\int d^3\bx
N\sqrt{g}}\Pi_N \ .
\end{equation}
In other words, we decompose the constraint $\pi_N(\bx)$ in terms of
the local and global constraints $\tpi_N(\bx)$ and $\Pi_N$.
The local constraint $\tpi_N(\bx)$ contains one
constraint per point in space minus one global constraint
\footnote{In the notation used in \cite{Kuchar:1991xd}, such a
constraint is said to contain a total of $\infty^{3}-1$ constraints.},
since these constraints are restricted by definition as
\begin{equation}
\int d^3\bx N(\bx)\tpi_N(\bx)=0 \ .
\end{equation}
Together with the global constraint $\Pi_N$ we have a total of
one constraint per point in space, which is the same as the number
of the original constraints $\pi_N$.

Now we are ready to proceed to  the analysis of the stability of the
primary constraints.  The requirement of the  preservation of the
constraints $p_\Lambda\approx 0 \ , p_A\approx 0$ implies following
secondary constraints
\begin{eqnarray}
\Phi_I &\equiv & \frac{1}{\kappa^2} \sqrt{g}({}^{(3)}R-\Omega+\eta_1 a_i
a^i
%+\eta_2 \nabla_i a^i
)\approx 0 \ \ ,
\nonumber \\
\Phi_{II}&\equiv&  2g^{ij}\mG_{ijkl}\pi^{kl}+(\gamma_0+\gamma_1 a_i
a^i
%+ \gamma_2 \nabla_i a^i
+2\gamma_3\Lambda -\frac{1}{2}
\Lambda)g^{ij}\mG_{ijkl}g^{kl}\approx 0 \ .
\end{eqnarray}
 As usual the preservation of the constraint $\pi_i\approx 0$
implies the constraint $\mH_i$ that we extend with the appropriate
combinations of the primary constraints $p_A\approx 0 \ ,
p_\Lambda\approx 0$
\begin{equation}
\tmH_i=-2g_{ik}\nabla_j \pi^{jk}+p_A\partial_i A+p_\Lambda
\partial_i \Lambda \ .
\end{equation}
It is convenient to introduce the smeared form of these constraints
\begin{equation}
\bT_S(N^i)=\int d^3\bx N^i\tmH_i \ .
\end{equation}
Then it is easy to see that
\begin{eqnarray}\label{pbbtSgp}
\pb{\bT_S(N^i),g_{ij}}&=&-N^k\partial_k g_{ij}-
\partial_i N^kg_{kj}-g_{ik}\partial_j N^k  \ , \nonumber \\
\pb{\bT_S(N^i),\pi^{ij}}&=&-\partial_k (N^k\pi^{ij}) +
\partial_k N^i\pi^{kj}+\pi^{ik}\partial_k N^j \ ,  \nonumber \\
\pb{\bT_S(N^i),\Lambda}&=&-N^i\partial_i \Lambda \ , \nonumber \\
\pb{\bT_S(N^i),A}&=&-N^i\partial_i A \ ,
\end{eqnarray}
so that
\begin{eqnarray}\label{pbbTSPhi}
\pb{\bT_S(N^i),\Phi_{II}}&=&-\partial_k (N^k\Phi_{II})\approx 0 \ ,
\nonumber \\
\pb{\bT_S(N^i),\Phi_I}&=&-\partial_k (N^k\Phi_I)\approx 0  \ .
\end{eqnarray}
Let us now consider the time evolution of the global constraint
$\Pi_N$
\begin{equation}
\partial_t \Pi_N=\pb{\Pi_N,\int d^3\bx N\mH_0}=
\int d^3\bx N\mH_0=0  \ ,
\end{equation}
where we also used the fact that
\begin{equation}
\pb{\Pi_N,\Phi_I(\bx) }=\pb{\Pi_N,\Phi_{II}(\bx)}=0 \ .
\end{equation}
We see that  the requirement of the preservation of the constraint
$\Pi_N\approx 0$ implies an existence of the second global
constraint
\begin{equation}
\Pi_0\equiv \int d^3\bx N\mH_0 \approx 0 \ .
\end{equation}
Finally, the requirement of the preservation of the constraint
$\tpi_N\approx 0$ implies
\begin{eqnarray}
\partial_t \tpi_N(\bx)&=&
\pb{\pi_N(\bx)-\frac{\sqrt{g}(\bx)}{\int d^3\bx N\sqrt{g}}\Pi_N,
\int d^3\by N\mH_0} \nonumber \\
&=&\mC(\bx)-\frac{\sqrt{g}(\bx)}{\int d^3\bx
N\sqrt{g}}\Pi_0\equiv \tmC(\bx) \ ,
\end{eqnarray}
where
\begin{eqnarray}\label{defmC}
\mC&=&\mH_0-\frac{1}{N}\nabla_i V^i -(\frac{2}{\kappa^2}a_m a^m
\Lambda \gamma_1+\frac{2}{\kappa^2}\gamma_1
\nabla_i[a^i\Lambda])\sqrt{g}g^{ij}\mG_{ijkl}g^{kl} \nonumber \\
&-&(\frac{2}{\kappa^2}a_m a^m A \eta_1+\frac{2}{\kappa^2}\eta_1
\nabla_i[a^i A])\sqrt{g} \ ,
\end{eqnarray}
and where we defined vector density
\begin{eqnarray}
V^i(\bx)=\frac{1}{\kappa^2}\frac{\delta }{\delta a_i(\bx)} \int
N\sqrt{g}\mV(g_{ij},a_i) \ , %\nonumber \\
\end{eqnarray}
and where $\mH_0$ is equal to
\begin{eqnarray}
\mH_0&=&\frac{\kappa^2}{\sqrt{g}} (\pi^{ij}-\frac{1}{2}\Lambda
\sqrt{g}g^{ij}) \mG_{ijkl} (\pi^{kl}-\frac{1}{2}\Lambda
\sqrt{g}g^{kl}) \nonumber \\
&+&\frac{1}{\kappa^2}\sqrt{g}\mV(g,a_i)- \frac{1}{\kappa^2}\sqrt{g}
( \gamma_0+\gamma_1 a_i a^i\Lambda
%+\gamma_2 \nabla_i
%a^i
+\gamma_3\Lambda^2)
g^{ij}\mG_{ijkl}g^{kl} \ .
\end{eqnarray}
Note that $\mC$ defined in (\ref{defmC}) is an extended version of
the constraint introduced in \cite{Donnelly:2011df}.

 Before we proceed further we show that with the help
of $\mC$ we can write $\int d^3\bx N\mH_0$ as
\begin{eqnarray}
\int d^3\bx N\mH_0 =\int d^3\bx N\mC \ ,
\end{eqnarray}
when we presume that the spatial hypersurface does not have a
boundary. Using this fact we obtain that $\tmC$ obeys following
condition
\begin{equation}
\int d^3\bx N\tmC=0 \ .
\end{equation}
Again we see that the local constraint $\tmC(\bx)$ together with
the global constraint $\Pi_0$ contain one constraint per each point in
space. Collecting all these constraints together we find that the
total Hamiltonian has the form
\begin{equation}\label{HTnew}
H_T=\Pi_N+\Pi_0+\int d^3\bx (v_N\tpi_N+v^i\pi_i+v^A p_A+v^\Lambda
p_\Lambda +\Gamma^I\Phi_I+\Gamma^{II}\Phi_{II}+\Gamma^{\tmC} \tmC) \
.
\end{equation}
Before we proceed further we list a collection of useful Poisson
brackets
\begin{eqnarray}\label{pAmC}
\pb{p_A(\bx),\mC(\by)}&=&\frac{2}{\kappa^2}a_m a^m\eta_1
\delta(\bx-\by)+\frac{2}{\kappa^2}\eta_1 \nabla_{y^i}[
a^i(\by)\delta(\bx-\by)] \ , \nonumber \\
\pb{p_\Lambda(\bx),\mC(\by)}&=&(\frac{2}{\kappa^2}a_m a^m\gamma_1
\delta(\bx-\by)+\frac{2}{\kappa^2}\gamma_1 \nabla_{y^i}[
a^i(\by)\delta(\bx-\by)])g^{ij}\mG_{ijkl}g^{kl}(\by) \  \nonumber \\
\end{eqnarray}
and also
\begin{equation}
\pb{p_\Lambda(\bx),\Phi_{II}(\by)}=-\left(2\gamma_3-\frac{1}{2}\right)g^
{ij}
\mG_{ijkl}g^{kl}(\bx)\delta(\bx-\by)
\end{equation}
and we see that this Poisson bracket is zero for
$\gamma_3=\frac{1}{4}$. It is also clear that
\begin{equation}
\pb{\tpi_N(\bx),\mC(\by)}=\triangle_{\pi_N,\mC}(\bx,\by)\neq 0 \
\end{equation}
and also
\begin{eqnarray}
\pb{\tpi_N(\bx),\Phi_I(\by)}&=&\frac{\eta_1}{\kappa^2}\frac{\sqrt{g}
(\by)}
{N(\by)}[a_i(\by)\delta(\bx-\by)
-\partial_{y^i}\delta(\bx-\by)]a^i(\by)\equiv
\triangle_{\pi_N,\Phi_I}(\bx,\by) \ , \nonumber \\
\pb{\tpi_N(\bx),\Phi_{II}(\by)}&=&\frac{\gamma_1}{\kappa^2}\frac{\sqrt{g
}(\by)}
{N(\by)}[a_i(\by)\delta(\bx-\by)
-\partial_{y^i}\delta(\bx-\by)]a^i(\by)g^{ij}\mG_{ijkl}g^{kl}\equiv
\triangle_{\pi_N,
\Phi_{II}}(\bx,\by) \ . \nonumber \\
\end{eqnarray}
 Observe that $\Pi_N$ has vanishing Poisson bracket with
$\mC$ as follows from the following Jacobi identity
\begin{eqnarray}
\pb{\Pi_N,\mC(\bx)}&=&\pb{\Pi_N,\pb{\pi_N(\bx),\int d^3\by N\mH_0}}
\nonumber \\
&=&-\pb{\pi_N(\bx),\pb{\int d^3\by N\mH_0,\Pi_N}}- \pb{\int d^3\by
N\mH_0,\pb{\Pi_N,\pi_N(\bx)}} \nonumber \\
&=&-\pb{\pi_N(\bx),\int d^3\by N\mH_0}-\pb{\int d^3\by
N\mH_0,\pi_N(\bx)}=0 \ ,
\end{eqnarray}
using the fact that $\pb{\Pi_N,\mH_0(\bx)}=0$. Then it is easy to
see that $\pb{\Pi_N,\tmC(\bx)}=0$ and also
\begin{equation}
\pb{\Pi_N,\Phi_I(\bx) }=\pb{\Pi_N,\Phi_{II}(\bx)}=0
\end{equation}
so that we could anticipate that $\Pi_N$ is the first class
constraint.

Next we will discuss the constraint structure of the theory
for specific values of the parameters $\gamma_i,\eta_i$.

\subsection{Generic case: $\gamma_3\neq \frac{1}{4},
\gamma_1\neq 0 \ , \eta_1\neq 0$} In this subsection we  denote  all
constraints as $\Psi_i=(p_\Lambda,\Phi_I,\tmC,\tpi_N,p_A,\Phi_{II})$
and corresponding Poisson brackets between these constraints as
\begin{equation}
\pb{\Psi_i(\bx),\Psi_j(\by)}=\triangle_{ij}(\bx,\by) \
\end{equation}
with inverse matrix $\triangle^{ij}(\bx,\by)$ that obeys the
equation
\begin{equation}\label{definverse}
\int d^3\bz \triangle_{ik}(\bx,\bz)\triangle^{kj}(\bz,\by)=
\delta_i^j \delta(\bx-\by) \ .
\end{equation}
Now we analyze the time evolution of all constraints. For $p_A$ we
have
\begin{equation}
\partial_t p_A(\bx)=\pb{p_A(\bx),H_T}=\int d^3\by
\triangle_{p_A,\tmC}(\bx,\by)\Gamma^{\tmC}(\by)=0  \
\end{equation}
that implies $\Gamma^{\tmC}=0$ as follows from (\ref{pAmC}).
 In the same way we get
\begin{equation}
\partial_t p_\Lambda(\bx)=\pb{p_\Lambda(\bx),H_T}=
\int d^3\by
\triangle_{p_\Lambda,\Phi_{II}}(\bx,\by)\Gamma^{II}(\by)=0 \ ,
\end{equation}
where we used the fact that $\Gamma^{\tmC}=0$. We again see that the
equation above has the solution $\Gamma^{II}=0$. Finally
\begin{equation}\label{piNI}
\partial_t\tpi_N(\bx)=\pb{\tpi_N(\bx),H_T}=
\int d^3\by \triangle_{\tpi_N,\Phi_I}(\bx,\by)\Gamma^{I}=0 \ ,
\end{equation}
where we used the fact that $\Gamma^{\tmC}=\Gamma^{II}=0$. Then the
equation (\ref{piNI}) implies $\Gamma^I=0$. Now using these
results it is easy to perform the analysis of the preservation of
the constraints $\Phi_I\approx 0 \ , \Phi_{II}\approx 0 $ and
$\tmC\approx 0$. However, we should also ensure that the constraints
$\Pi_N,\Pi_0$ are the first class constraints. To do this we
introduce following modification of these constraints
\begin{eqnarray}
\tilde{\Pi}_N&=&\Pi_N-\int d^3\bz d^3\bz'\pb{\Pi_N,\Psi_i(\bz)}
\triangle^{ij}(\bz,\bz')\Psi_j(\bz') \ , \nonumber
\\
\tilde{\Pi}_0&=&\Pi_0-\int d^3\bz d^3\bz'\pb{\Pi_0,\Psi_i(\bz)}
\triangle^{ij}(\bz,\bz')\Psi_j(\bz') \ ,
\end{eqnarray}
which by definition Poisson commute with all second class
constraints
$\Psi_i$ as can be seen from
\begin{eqnarray}\label{PiNpb}
\pb{\tilde{\Pi}_N,\Psi_i(\bx)}&=&\pb{\Pi_N,\Psi_i(\bx)}- \int d^3\bz
d^3\bz'\pb{\Pi_N,\Psi_k(\bz)}
\triangle^{kj}(\bz,\bz')\pb{\Psi_j(\bz'),\Psi_i(\bx)}
 \nonumber \\
&=&\pb{\Pi_N,\Psi_i(\bx)}- \int d^3\bz d^3\bz'\pb{\Pi_N,\Psi_k(\bz)}
\triangle^{kj}(\bz,\bz')\triangle_{ji}(\bz',\bx)
 \nonumber \\
 &=&\pb{\Pi_N,\Psi_i(\bx)}- \int d^3\bz \pb{\Pi_N,\Psi_k(\bz)}
\delta(\bz-\bx)\delta^k_i=0 \ . \nonumber \\
\end{eqnarray}
In the same way we find $\pb{\tilde{\Pi}_0,\Psi_i(\bx)}=0$. We see
that it is natural to replace $\Pi_N,\Pi_0$ with $\tilde{\Pi}_N$ and
$\tilde{\Pi}_0$ that are the first class constraints and Poisson
commute with $\Psi_i$. Then we can easily perform the analysis of
the time evolution of the constraints $\Phi_{I},\Phi_{II}$ and
$\tmC$ where now the requirement of their preservations imply that
$v_N,v_A$ and $v_\Lambda$ have to vanish.

In summary we have the collection of the second class constraints
$\Psi_i$ that can be solved in the following way. From $\Phi_{II}$
we express $\Lambda$ as function of canonical variables. From $\tmC$
we express $A$ at least in principle and from $\Phi_I$ we express
$a_i$ that allows us to find $N$ again at least in principle. In
other words all phase space variables $(N,\pi_N),(A,p_A)$ and
$(\Lambda,p_\Lambda)$ are eliminated. On the other hand there are
still $12$ degrees of freedom in $g_{ij},\pi_{ij}$ where $6$ of them
can be eliminated by gauge fixing of three  first class constraints
$\tmH_i$. In other words the generic case has an important property
that the scalar graviton is still present. Finally we have two
global first class constraints
$\tilde{\Pi}_N=\Pi_N,\tilde{\Pi}_0=\Pi_0$ where we used the fact
that the second class constraints $\Psi_i$ vanish strongly.

%%%%%%%%%%%%%%%%%%%%%%%%
\subsection{The Case $\gamma_3=\frac{1}{4}$}
In this case we find that $\pb{p_\Lambda(\bx),\Phi_{II}(\by)}=0$ and
also that $\Phi_{II}$ does not depend on $\Lambda$. Now we proceed
in the following way. Let us denote $\Psi_i$ as collection of the
constraints $(\tmC,\tpi_N,p_A,\Phi_{I})$ and the matrix of Poisson
brackets between them as
\begin{equation}\label{deftrianglegamma}
\pb{\Psi_i(\bx),\Psi_j(\by)}=\triangle_{ij}(\bx,\by) \ .
\end{equation}
It is important to stress that there are still non-zero Poisson
brackets between $p_\Lambda$ and $\Psi_i$ and $\Phi_{II}$.
 Then we define following constraint
\begin{eqnarray}\label{deftildePhiII}
\tilde{\Phi}_{II}(\bx)=\Phi_{II}(\bx)- \int d^3\by d^3\bz
\pb{\Phi_{II}(\bx),\Psi_i(\by)}\triangle^{ij}(\by,\bz)\Psi_j(\bz) \ .
\end{eqnarray}
It can be shown as in (\ref{PiNpb}) that this constraint Poisson
commutes with all second class constraints $\Psi_i$
\begin{eqnarray}
\pb{\tilde{\Phi}_{II}(\bx),\Psi_k(\by)}&=&
%\pb{\Phi_{II}(\bx),\Psi_k(\by)} \nonumber \\
%-\int d^3\bz
%d^3\bz'\pb{\Phi_{II}(\bx),\Psi_i(\bz)}\triangle^{ij}(\bz,\bz')\pb{
% \Psi_j(\bz'),\Psi_k(\by)}=
%\nonumber \\
\pb{\Phi_{II}(\bx),\Psi_k(\by)} \nonumber \\
&-&\int d^3\bz d^3\bz'\pb{\Phi_{II}(\bx),\Psi_i(\bz)}
\triangle^{ij}(\bz,\bz')\triangle_{jk}(\bz',\by)=0 \ ,
%\nonumber \\
%&=&\pb{\Phi_{II}(\bx),\Psi_k(\by)}-\int d^3\bz
%\pb{\Phi_{II}(\bx),\Psi_i(\bz)}\delta^i_k\delta(\bz-\by)=0 \ ,
\end{eqnarray}
where the matrix $\triangle^{ij}(\bx,\by)$ is the inverse matrix to
the matrix $\triangle_{ij}(\bx,\by)$ defined in
(\ref{deftrianglegamma}) that has the property
\begin{equation}
\int d^3\bz\triangle_{ij}(\bx,\bz)\triangle^{jk}(\bz,\by)=
\delta_i^k\delta(\bx-\by) \ .
\end{equation}
In the same way we define $\tilde{p}_\Lambda$ as
\begin{equation}
\tilde{p}_\Lambda(\bx)=p_\Lambda(\bx)- \int d^3\bz
d^3\bz'\pb{p_\Lambda(\bx),\Psi_i(\bz)}\triangle^{ij}(\bz,
\bz')\Psi_j(\bz') \ ,
\end{equation}
which again obeys
\begin{equation}
\pb{\tilde{p}_\Lambda(\bx),\Psi_i(\by)}=0 \ .
\end{equation}
On the other hand  the Poisson bracket between $\tilde{p}_\Lambda$
and $\tilde{\Phi}_{II}$ is equal to
\begin{equation}
\pb{\tilde{p}_\Lambda(\bx),\tilde{\Phi}_{II}(\by)}=- \int d^3\bz
d^3\bz'\pb{p_\Lambda(\bx),\Psi_i(\bz')}\triangle^{ij}(\bz,\bz')
\pb{\Psi_j(\bz'),\Phi_{II}(\by)}
\end{equation}
that is non-zero and hence we see that  $\tilde{p}_\Lambda(\bx),
\tilde{\Phi}_{II}(\by)$ are the second class constraints. Finally we
define $\tilde{\Pi}_N,\tilde{\Pi}_0$ as
\begin{eqnarray}\label{deftildePiA}
\tilde{\Pi}_N&=&\Pi_N-\int d^3\bz d^3\bz'\pb{\Pi_N,\Psi_A(\bz)}
\triangle^{AB}(\bz,\bz')\Psi_B(\bz') \ , \nonumber
\\
\tilde{\Pi}_0&=&\Pi_0-\int d^3\bz d^3\bz'\pb{\Pi_0,\Psi_A(\bz)}
\triangle^{AB}(\bz,\bz')\Psi_B(\bz') \ ,
\end{eqnarray}
where $\Psi_A=(\tilde{p}_\Lambda,\tilde{\Phi}_{II},\Psi_i)$ and
where the matrix $\triangle_{AB}$ is the matrix of the Poisson
brackets between these constraints that has inverse $\triangle^{AB}$
by definition. Using this notation we find the total Hamiltonian in
the form
\begin{equation}
H_T=\tilde{\Pi}_0+c_N\tilde{\Pi}_N +\int d^3\bx (\Gamma^A \Psi_A
+v^i\pi_i+N^i\tmH_i)  \ .
\end{equation}
 Now we are ready to study the time evolution of all
constraints
\begin{eqnarray}
\partial_t \tilde{p}_\Lambda(\bx)=
\pb{\tilde{p}_\Lambda(\bx),H_T}= \int d^3\by
\pb{\tilde{p}_\Lambda(\bx),
\tilde{\Phi}_{II}(\by)}\Gamma^{II}(\by)=0 \ ,
\end{eqnarray}
that has solution $\Gamma^{II}=0$. Then it is easy to see that
\begin{equation}
\partial_t \tilde{\Phi}_{II}(\by)=
\pb{\tilde{\Phi}_{II}(\by),H_T}= \int d^3\by
\pb{\tilde{\Phi}_{II}(\bx),
\tilde{p}_\Lambda(\by)}v_\Lambda(\by)=0 \   \nonumber \\
\end{equation}
that has again solution $v_\Lambda=0$.  Then we can proceed to the
analysis of the time evolution of the constraints $\Psi_i$. In case
of $p_A$ and $\pi_N$ we obtain
\begin{equation}
\partial_t p_A(\bx)=\pb{p_A(\bx),H_T}\approx
\int d^3\by \pb{p_A(\bx),\tmC(\by)}\Gamma^{\tmC}(\by)=0 \ ,
\end{equation}
which gives $\Gamma^{\tmC}=0$. In the same way we have
\begin{equation}
\partial_t \tpi_N(\bx)=\pb{\tpi_N(\bx),H_T}\approx
\int d^3\by \pb{\tpi_N(\bx),\Phi_I(\by)}\Gamma^I(\by)=0 \ ,
\end{equation}
which again implies $\Gamma^I$. Using these results it is easy to
perform the analysis of the time evolution of the constraints $\tmC$
and $\Phi_I$. We again find two  equations for the Lagrange
multipliers $v^N, v^A$ that can be solved for the canonical
variables. In summary, we have six second class constraints
$\tilde{\Phi}_{II},\tilde{p}_\Lambda,\tmC,\tpi_N,\Phi_I,p_A$ that
can be solved in the same way as in previous section. In other
words, the scalar graviton is still present.
%%%%%%%%%%%%%%%%%%%%%%%%%%
\subsection{The Case: $\gamma_3=\frac{1}{4}\ , \gamma_1=0$}
This is an exceptional case  when
\begin{equation}
\pb{p_\Lambda(\bx),\Phi_{II}(\by)}=0 \ , \quad
\pb{p_\Lambda(\bx),\tmC(\by)}=0 \ .
\end{equation}
 In other words $p_\Lambda\approx
0$ is the first class constraint. On the other hand
$\Phi_{II}\approx 0$ has still non-zero Poisson brackets with
$\tmC\approx 0$ and with $\Phi_I\approx 0$. As in previous section
we use the common notation $\Psi_i=(\tmC,\tpi_N,p_A,\Phi_I)$ and
introduce $\tilde{\Phi}_{II}$ as in (\ref{deftildePhiII}).  Now we
are ready to study the time evolution of all constraints. In case
$p_\Lambda$ the situation is trivial as $p_\Lambda$ is the first
class constraint. In case of the constraints
$\tilde{\Phi}_{II}\approx 0$ we obtain
\begin{eqnarray}
\partial_t \tilde{\Phi}_{II}(\bx)=\pb{\tilde{\Phi}_{II}(\bx),H_T}
=c_N(t)\int  \pb{\tilde{\Phi}_{II}(\bx),\mH_0(\by)}N(\by)\equiv
c_N(t)\Phi_{III}(\bx)=0 \ .  \nonumber \\
\end{eqnarray}
In other words, the requirement of the preservation of the
constraint $\tilde{\Phi}_{II}\approx 0$ either imposes the condition
$c_N(t)=0$ or we should introduce another local constraint
$\Phi_{III}\approx 0$. Since $\Phi_{II}$ is local constraint we mean
that it is more natural to impose another local constraint rather
then to determine global Lagrange multiplier to be zero. In other
words we claim that the requirement of the preservation of the
constraint $\tilde{\Phi}_{II}\approx 0$ induces another constraint
$\Phi_{III}\approx 0$ \footnote{ This is similar situation as in
paper \cite{Bellorin:2013zbp}.}. Now we proceed in the similar way
as in previous section. Let us denote all second class constraints
as $\Psi_A=(\tilde{\Phi}_{II},\Phi_{III},\Psi_i)$ and introduce
$\tilde{\Pi}_N,\tilde{\Pi}_0$ as in (\ref{deftildePiA}) that ensure
that $\tilde{\Pi}_N,\tilde{\Pi}_0$ are global first class
constraints. On the other hand  the existence of the constraints
$\Phi_I,p_A$ does not restrict the number of the physical degrees of
freedom in the gravity sector since we again have non-zero Poisson
brackets between $\tmC$ and $p_A$  and $\Phi_I$ and $\tpi_N$ due to
the presence of the term $\eta_1 a_i a^i$ in $\Phi_I$.
 Explicitly, the time evolution of the constraint
$p_A$ is given by the equation
\begin{equation}
\partial_t p_A(\bx)=\pb{p_A(\bx),H_T}=\int d^3\bx
\pb{p_A(\bx),\tmC(\by)}\Gamma^{\tmC}(\by)=0 \ ,
\end{equation}
which again implies $\Gamma^{\tmC}=0$. In the same way the time
evolution of the constraint $\tpi_N\approx 0$ implies that
$\Gamma^I=0$. Finally the requirement of the preservation of the
constraint $\tmC\approx 0, \Phi_I\approx 0$ implies that  $u_N,u_A$
are zero. In other words $\Phi_I,\tmC,\tpi_N,p_A$ are the second
class constraints, where $\tmC$ can be solved for $A$ while $\Phi_I$
can be solved for $a_i$ and hence for $N$, at least in principle.
Further, $p_\Lambda\approx 0$ is the first class constraint that can
be fixed by requirement $\Lambda=\mathrm{const}$.  However, as
opposite to the previous cases we now have two constraints
$\tilde{\Phi}_{II}$ and $\Phi_{III}$ that are the second class
constraints that can be solved for two degrees of freedom that are
contained in $g_{ij}$. For example, from $\tilde{\Phi}_{II}$ we can
express $\pi=\pi^{ij}g_{ij}$ at least in principle. In summary, the
exceptional case when $\gamma_1=0,\gamma_3=\frac{1}{4}$ allows to
eliminate the scalar graviton. However, now due to the fact that
$\Phi_{III}$ arises from the Poisson bracket between
$\tilde{\Phi}_{II}$ and $\mH_0$ we find that $\Phi_{III}$ is very
complicated expression in the canonical variables. Further, the
symplectic structure of this case is complicated as well due to
the non-trivial form of the Poisson brackets between all second
class constraints.

\subsection{Exceptional case: $\gamma_1=\gamma_2=\eta_1=\eta_2=0$}
In this subsection we consider exceptional case when
$\gamma_1=\gamma_2=\eta_1=\eta_2=0$ and $\gamma_3=\frac{1}{4}$. Note
that in this case the constraints $\Phi_{I},\Phi_{II}$ and $\tmC$
have very simple form
\begin{eqnarray}\label{PhiIIIex}
\Phi_{II}&=& 2g^{ij}\mG_{ijkl}\pi^{kl}+\gamma_0
g^{ij}\mG_{ijkl}g^{kl}\approx 0 \ ,
\nonumber \\
\Phi_I&=& \frac{1}{\kappa^2} \sqrt{g}({}^{(3)}R-\Omega)\approx 0 \ ,
\nonumber \\
\tmC&=&\mH_0-\frac{1}{N}\nabla_i V^i-\frac{\sqrt{g}}{\int d^3\bx
N\sqrt{g}}\Pi_0  \ ,
\end{eqnarray}
where now $\tmC$ does not depend on $\Lambda$ and $A$. Now we see
that $p_A$ and $p_\Lambda$ have vanishing Poisson brackets with all
other constraints so that they are the first class constraints. We
also see that we have
\begin{eqnarray}
\pb{\tpi_N(\bx),\Phi_{I}(\by)}=\pb{\tpi_N(\bx),\Phi_{II}(\by)}=0
\end{eqnarray}
Let us denote $\Psi_i=(\tpi_N,\tmC)$ and corresponding matrix of
Poisson brackets as
$\pb{\Psi_i(\bx),\Psi_j(\by)}=\triangle_{ij}(\bx,\by)$ (note that
following Poisson bracket is non-zero as well
$\pb{\tmC(\bx),\tmC(\by)}$) with inverse $\triangle^{ij}$. We again
define $\tilde{\Phi}_I,\tilde{\Phi}_{II}$ as
\begin{eqnarray}
\tilde{\Phi}_I(\bx)=\Phi_I(\bx)-\int d^3\bz d^3\bz'
\pb{\Phi_I(\bx),\Psi_i(\bz)}\triangle^{ij}(\bz,\bz' )\Psi_j(\bz')
\ , \nonumber \\
\tilde{\Phi}_{II}(\bx)=\Phi_{II}(\bx)-\int d^3\bz d^3\bz'
\pb{\Phi_I(\bx),\Psi_i(\bz)}\triangle^{ij}(\bz,\bz' )\Psi_j(\bz')
\ .
\end{eqnarray}
Then it is easy to see that
\begin{eqnarray}
\pb{\tilde{\Phi}_I(\bx),\Psi_i(\by)}&=&\pb{\Phi_I(\bx),\Psi_i(\by)}-
\int d^3\bx \pb{\Phi_I(\bx),\Psi_k(\bz)}\triangle^{kl}(\bz,\bz')
\triangle_{li}(\bz',\by)
\nonumber \\
&=&\pb{\Phi_I(\bx),\Psi_i(\by)}-\int d^3\bz \pb{\Phi_I(\bx),\Psi_k(\bz)}
\delta^k_i\delta(\bz-\by)=0 \ .  \nonumber \\
%&=&\pb{\Phi_I(\bx),\Psi_i(\by)}- \pb{\Phi_I(\bx),\Psi_i(\by)}
%=0 \ .
\end{eqnarray}
In the same way we find that
$\pb{\tilde{\Phi}_{II}(\bx),\Psi_i(\by)}=0$. On the other hand
we clearly have that there is a non-zero matrix
\begin{equation}
\pb{\tilde{\Phi}_A(\bx),\tilde{\Phi}_B(\by)}=\Omega_{AB}(\bx,\by) \ ,
\end{equation}
where now $A,B=I,II$.

Now we are ready to proceed to the analysis of the time development
of various constraints. First of all we introduce
$\tilde{\Pi}_N,\tilde{\Pi}_0$ as in (\ref{deftildePiA}) where now
$\Psi_A=(\tilde{\Phi}_I,\tilde{\Phi}_{II},\tmC,\tpi_N,p_A,p_\Lambda)$.
 In case of the constraints $p_A$ and
$p_\Lambda$ we trivially obtain that they are preserved during the
time evolution of the system. In case of the constraints
$\tilde{\Phi}_A$ we obtain
\begin{eqnarray}
\partial_t \tilde{\Phi}_A(\bx)=\pb{\tilde{\Phi}_A(\bx),H_T}=
\int d^3\by  \Omega_{AB}(\bx,\by)\Gamma^B(\by)=0 \ ,
\end{eqnarray}
which can be solved for $\Gamma_B$ thanks to the fact that the
matrix $\Omega_{AB}$ is
non-singular. In case of the constraints $\Psi_i$ we find
\begin{eqnarray}
\partial_t\Psi_i(\bx)=\pb{\Psi_i(\bx),H_T}=
\int d^3\by \triangle_{ij}(\bx,\by) \Gamma^j(\by)=0 \ ,
\end{eqnarray}
which can be again solved for $\Gamma^i$. In other words we have
completely fixed all Lagrange multipliers. Now we have following
picture.
 The constraints $\tmC\approx 0 \ , \tpi_N\approx 0$ are
the second class constraints that can be solved for $N$ and $\pi_N$.
On the other hand the constraints $\tilde{\Phi}_I\approx 0 \ ,
\tilde{\Phi}_{II}\approx 0$ are the second class constraints that
can be solved for two modes corresponding to the scalar graviton.
Explicitly, from $\Phi_{II}$ given in (\ref{PhiIIIex}) we can easily
express $\pi=g^{ij}\pi_{ij}$ as constant. On the other hand from
$\Phi_{I}$ we could express another mode.  Note that the structure
of these constraints is much simpler than in previous section that
makes this exceptional case more attractive.  In summary, we have
found the non-projectable HL gravity with the physical spectrum that
is the same as in General Relativity.

\section{Projectable HL gravity with additional
constraints}\label{fourth}
 In this section we present the
Hamiltonian analysis of projectable version of HL gravity with
additional constraints. Recall that in this case the action has the
form
\begin{eqnarray}\label{Sproj}
S&=&\frac{1}{\kappa^2} \int dt d^3\bx [N\sqrt{g}(K_{ij}
\mG^{ijkl}K_{kl}-\mV(g)+\Lambda \sqrt{g}K \nonumber \\
&+&\frac{1}{4}\sqrt{g}f(\Lambda)g^{ij}\mG_{ijkl}g^{kl}
+A\sqrt{g}({}^{(3)}R-\Omega)] \ ,
\end{eqnarray}
where now $N=N(t)$ and where the potential $\mV$ has the same form
as in non-projectable case with exception that all terms that
contain $a_i$ are missing. Finally the function $f(\Lambda)$ has the
form
\begin{equation}
f(\Lambda)=\gamma_0\Lambda+\gamma_3\Lambda^2 \ .
\end{equation}
Now we can proceed to the Hamiltonian formulation of the
projectable theory \eqref{Sproj}.
If we proceed in the same way as in section~\ref{third} we find
the bare Hamiltonian in the form
\begin{equation}
H_B = \int  d^3\bx [N\mH_0 - \frac{1}{\kappa^2}
 AN\sqrt{g}({}^{(3)}R-\Omega)+N^i\mH_i] \ ,
\end{equation}
where
\begin{eqnarray}
\mH_0&=&\frac{\kappa^2}{\sqrt{g}} (\pi^{ij}-\frac{1}{2}\Lambda
\sqrt{g}g^{ij}) \mG_{ijkl} (\pi^{kl}-\frac{1}{2}\Lambda
\sqrt{g}g^{kl})  \nonumber \\
&+&\frac{1}{\kappa^2}\sqrt{g}\mV(g)
-\frac{1}{\kappa^2}\sqrt{g}(\gamma_0\Lambda
+\gamma_3\Lambda^2)g^{ij}\mG_{ijkl}g^{kl} \ .
\end{eqnarray}
Note that there is also collection of local primary constraints
\begin{equation}
 \pi_i(\bx)\approx 0 \ , \quad p_A(\bx)\approx 0 \ ,
\quad p_\Lambda(\bx) \approx 0 \
\end{equation}
together with the global one
\begin{equation}
\pi_N\approx 0 \ \ .
\end{equation}
 Now we proceed to the analysis of the
stability of the primary constraints.  The requirement of the
preservation of the constraints $p_\Lambda\approx 0 \ , p_A\approx
0$ implies following secondary constraints
\begin{eqnarray}
\Phi_I &\equiv & \frac{1}{\kappa^2}
\sqrt{g}({}^{(3)}R-\Omega)\approx 0 \ \ ,
\nonumber \\
\Phi_{II}&\equiv&
2g^{ij}\mG_{ijkl}\pi^{kl}+(\gamma_0+2\gamma_3\Lambda -\frac{1}{2}
\Lambda)g^{ij}\mG_{ijkl}g^{kl}\approx 0 \ .
\end{eqnarray}
 As usual the preservation of the constraint $\pi_i\approx 0$
implies the constraint $\mH_i$ that we extend with the appropriate
combinations of the primary constraints $p_A\approx 0 \ ,
p_\Lambda\approx 0$
\begin{equation}
\tmH_i=-2g_{ik}\nabla_j \pi^{jk}+p_A\partial_i A+p_\Lambda
\partial_i \Lambda \ .
\end{equation}
It can be shown as in section (\ref{third}) that they are the first
class constraints that are generators of spatial diffeomorphism.
Finally the requirement of the preservation of the constraint
$\pi_N\approx 0$ implies following global constraint
\begin{equation}
\partial_t \pi_N(t)=\pb{\pi_N(t),H_B}=-\int d^3\bx \mH_0\equiv
-\Pi_0 \ .
\end{equation}
Then the total Hamiltonian with all constraints included has the
form
\begin{equation}
H_T=N(t)\Pi_0+v_N\pi_N+ \int d^3\bx (v^i\pi_i+v^A p_A+ v^\Lambda
p_\Lambda+\Gamma^I\Phi_I+\Gamma^{II}\Phi_{II}) \ .
\end{equation}
Now the further analysis depends on the value of the parameter
$\gamma_3$.
\subsection{The Case $\gamma_3\neq \frac{1}{4}$}
In this case we find that
\begin{equation}
\pb{p_\Lambda(\bx),\Phi_{II}(\by)}=-2(\gamma_3-\frac{1}{4})g^{ij}
\mG_{ijkl}g^{kl}\delta(\bx-\by) \ .
\end{equation}
We see that the constraints $p_\Lambda\approx 0$ and
$\Phi_{II}\approx 0$ are the second class constraints. Let us
further define  the modified constraint $\tPhi_{I}$ as
\begin{equation}
\tPhi_I(\bx)=\Phi_I(\bx)-\int d^3\bz
d^3\bz'\pb{\Phi_I(\bx),\Psi_{A}(\bz')}\triangle^{AB}(\bz,
\bz')\Psi_B(\bz') \ ,
\end{equation}
where $\Psi_A\equiv (p_\Lambda,\Phi_{II})$ and where
$\pb{\Psi_A(\bx),\Psi_B(\by)}=\triangle_{AB}(\bx,\by)$ with inverse
matrix $\triangle^{AB}$. Then we have that
$\pb{\tPhi_I(\bx),\Psi_{A}(\by)}=0$ and hence the time evolution of
the constraint $\tPhi_I$ is equal to
\begin{equation}
\partial_t\tPhi_I(\bx)=\pb{\tPhi_I,H_T}=N(t)\int
d^3\by\pb{\tPhi_I(\bx),\mH_0(\by)}\equiv N(t)\Phi_{III}=0 \ .
\end{equation}
Now we can argue as in previous section that the requirement of the
preservation of the constraint $\tPhi_I$ implies an additional
constraint $\Phi_{III}$ whose explicit form is not needed for us.
Then we have following collection of the second class constraints
$\Psi_i=(\tPhi_I,\Phi_{III},\Phi_{II},p_\Lambda)$ where the last two
constraints can be solved for $p_\Lambda$ and for $\Lambda$ while
$\tPhi_I$ and $\Phi_{III}$ can be solved for the scalar graviton in
the similar way as in \cite{Kluson:2011xx}.
%%%%%%%%%%%%%%%%%%%%%%%
\subsection{The Case $\gamma_1= \frac{1}{4}$}
In this case we have that $\pb{p_\Lambda(\bx),\Phi_{II}(\by)}=0$ and
hence $p_\Lambda\approx 0$ is the first class constraint. Then
$\Phi_A, A=I,II$ are the second class constraints with non-trivial
Poisson bracket
\begin{eqnarray}\label{triangleIII}
\pb{\Phi_{I}(\bx),\Phi_{II}(\by)}=
-\frac{2}{\kappa^2}\sqrt{g}
(\Lambda\delta(\bx-\by)+\nabla_i\nabla^i\delta(\bx-\by))
\equiv \triangle_{I,II} (\bx,\by) \ ,
\end{eqnarray}
using
\begin{equation}
\pb{R(\bx),\pi^{ij}(\by)}= -R^{ij}(\bx)\delta(\bx-\by)+\nabla^i
\nabla^j \delta(\bx-\by)-g^{ij} \nabla_k \nabla^k\delta(\bx-\by) \ .
\end{equation}
To proceed further we introduce modified form of the global
constraint in the form
\begin{equation}
\tilde{\Pi}_0=\Pi_0-\int d^3\bx \pb{\Pi_0,\Phi_A(\bx)}
\triangle^{AB}(\bx,\by)\Phi_B(\by) \ ,
\end{equation}
so that $\pb{\tilde{\Pi}_0,\Phi_A(\bx)}=0$. Then the   total
Hamiltonian has the form
\begin{equation}
H_T= N(t)\tilde{\Pi}_0+v_N\pi_N+ \int d^3\bx (v^A p_A+ v^\Lambda
p_\Lambda+\Gamma^I\Phi_I+\Gamma^{II}\Phi_{II}) \ .
\end{equation}
up to the diffeomorphism constraints. Then it is easy to study the
evolution of the constraints $\Phi_A$
\begin{eqnarray}
\partial_t\Phi_A(\bx)=\pb{\Phi_A,H_T}\approx \int d^3\by
\triangle_{AB}(\bx,\by)\Gamma^B(\by)=0 \ ,
\end{eqnarray}
which due to the fact that the matrix $\triangle_{AB}(\bx,\by)$ is
non-singular has the solution $\Gamma^B=0$. In other words
$\Phi_I,\Phi_{II}$ are the second class constraints. Further, the
constraints $p_\Lambda\approx 0 \ , p_A\approx 0$ are the first
class constraints  that can be fixed by particular choice of
$\Lambda$ and $A$. Further, from $\Phi_{II}$ we can eliminate the
trace of the conjugate momenta $g^{ij}\pi_{ij}$ while from $\Phi_I$
we can eliminate another degree of freedom from the graviton, at
least in principle.
 In summary the number of physical degrees of
freedom is the same as in case of General Relativity while the
symplectic structure is more involved. To see explicitly, let us
introduce  $G(\bx,\by)$ as  Green function of the operator
$\nabla_i\nabla^i+\Lambda$ defined as
\begin{equation}
\sqrt{g}(\nabla_i \nabla^i+\Lambda)G(\bx,\by)=\delta(\bx-\by) \ .
\end{equation}
Then it is easy to see that the inverse matrix to $\triangle_{I,II}$
has the form
\begin{equation}
(\triangle^{-1})^{II,I}(\bx,\by)=-\frac{\kappa^2}{2}G(\bx,\by) \ .
\end{equation}
Then we are ready to the calculations of the Dirac brackets of the
canonical variables
\begin{eqnarray}\label{sympro1}
& &\pb{g_{ij}(\bx),\pi^{kl}(\by)}_D=
\pb{g_{ij}(\bx),\pi^{kl}(\by)} \nonumber \\
&-& \int d^3\bz d^3\bz' \pb{g_{ij}(\bx),\Phi_{II}(\bz)}
(\triangle^{-1})^{II,I}(\bz,\bz')\pb{\Phi_I(\bz'),\pi^{kl}(\by)}
\nonumber \\
%\nonumber \\
%=\frac{1}{2}
% (\delta_i^k\delta_j^l+\delta_i^l\delta_j^k)\delta(\bx-\by)+\nonumber
%\\
%+ \frac{\kappa^2}{2}g_{ij}(\bx) \int d^3\bz' G(\bx,\bz')
%\frac{1}{\kappa^2}\sqrt{g}(\bz')
%(-R^{kl}(\bz')\delta(\bz'-\by)+\nonumber
%\\
%+\nabla^k\nabla^l_{z'}\delta(\bz'-\by)-g^{kl}(\bz')
%\nabla_{m,z'}\nabla^m \delta(\bz'-\by)) \nonumber \\
&=&\frac{1}{2}
(\delta_i^k\delta_j^l+\delta_i^l\delta_j^k)\delta(\bx-\by)-
\frac{1}{2}g_{ij}(\bx)G(\bx,\by)R^{kl}(\by) \nonumber \\
&+&\frac{1}{2}g_{ij}(\bx)\nabla_{\by}^k \nabla_{\by}^l
G(\bx,\by)\sqrt{g}(\by)
-\frac{1}{2}g_{ij}(\bx)\nabla_{m,\by}\nabla^m_{\by}G(\bx,\by)
g^{kl}(\by)\sqrt{g}(\by) \ .
%  \nonumber \\
\end{eqnarray}
In the same way we obtain
\begin{eqnarray}\label{sympro2}
& &\pb{\pi^{ij}(\bx),\pi^{kl}(\by)}_D= -\int d^3\bz
d^3\bz'\pb{\pi^{ij}(\bx),\Phi_I(\bz)}(\triangle^{-1})^{I,II}(\bz,\bz')
\pb{\Phi_{II}(\bz'),\pi^{kl}(\by)} \nonumber \\
&-&\int d^3\bz
d^3\bz'\pb{\pi^{ij}(\bx),\Phi_{II}(\bz)}(\triangle^{-1})^{II,I}(\bz,
\bz')
\pb{\Phi_{I}(\bz'),\pi^{kl}(\by)} \nonumber \\
&=&\frac{\kappa^2}{2}(R^{ij}(\bx)G(\bx,\by)\pi^{kl}(\by)-
\pi^{ij}(\bx)G(\bx,\by)R^{kl}(\by)) \nonumber \\
&-&\frac{\kappa^2}{2}\nabla^i_{\bx}\nabla^j_{\bx}
G(\bx,\by)\pi^{kl}(\by)+\frac{\kappa^2}{2}g^{ij}(\bx)\nabla_{k,\bx}
\nabla^k_{\bx}G(\bx,\by)\pi^{kl}(\by) \nonumber \\
&+&\frac{\kappa^2}{2}\pi^{ij}(\bx) \nabla^k_{\by} \nabla^l_{\by}
G(\bx,\by)-\frac{\kappa^2}{2}\pi^{ij}(\bx)g^{kl}(\by) \nabla_{k,\by}
\nabla^k_{\by} G(\bx,\by)
% \nonumber \\
\end{eqnarray}
using
\begin{eqnarray}
\pb{\pi^{ij}(\bx),\Phi_I(\by)}&\approx&
R^{ij}(\bx)\delta(\bx-\by)-\nabla^i_{\by} \nabla^j_{\by}
\delta(\bx-\by)+ g^{ij}\nabla_{k,\by} \nabla^k_{\by} \delta(\bx-\by)
\ ,
\nonumber \\
\pb{\Phi_{II}(\bx),\pi^{kl}(\by)}&=&\pi^{kl}(\bx)\delta(\bx-\by) \ ,
% \nonumber \\
\end{eqnarray}
and where $\nabla_{k,\by}$ means covariant derivative evaluated at
the point $\by$. We see from (\ref{sympro1}) and (\ref{sympro2})
that the symplectic structure of the theory \eqref{Sproj} is much
more complicated than in case of General Relativity. On the other hand
this symplectic structure and form of the constraints is much
simpler than in the generic case with $\gamma_3\neq \frac{1}{4}$,
which is the main reason why we introduced an additional constraint
in the projectable version of HL gravity. In any case the
projectable HL gravity with additional constraints has remarkable
property that the scalar graviton is eliminated.
\section{Conclusion}\label{fifth}
In this paper we performed the analysis of projectable and
non-projectable HL gravity with two additional constraints. We
showed that the structure of the constraints is more involved in the
case of non-projectable theory, since the form and number of
constraints depends on the values of the additional coupling
constants. We have shown that the scalar graviton is absent when the
coupling constants have the values $\gamma_3=\frac{1}{4},\gamma_1=0$
or when $\gamma_1=\eta_1=0$ and $\gamma_3=\frac{1}{4}$. In those
cases, the number of physical degrees of freedom is the same as in
GR. However, it is an open question whether these points are stable
under quantum corrections. In other words, even if we construct the
classical non-projectable HL gravity with the exceptional values of
the parameters given above, it is not clear whether quantum
corrections generate these operators.

In case of projectable theory the situation is different. We have
shown that in this case the scalar graviton is absent as well even
if the structure of the constraints depends on the value of the
parameter $\gamma_3$. However, the important point is that the
projectability condition is consistent truncation of the theory and
for that reason is expected to be stable under radiative
corrections. Hence we can conclude that the number of gravitational
degrees of freedom in the projectable HL gravity with additional
constraints is the same as in GR, despite of the fact that the
symplectic structures of the theories are different.

 \noindent {\bf Acknowledgements}
\\
The work of J.K. was supported by the Grant Agency of the Czech Republic
under the grant P201/12/G028. M.O. gratefully acknowledges support
from the Emil Aaltonen Foundation.

%%%%%%%%%%%%%%%%%%%%%%%%%%%%%%%%%%%%%%
%%%%%%% Thebibligraphy %%%%%%%%%%%
%%%%%%%%%%
%%%%%%%%%%%%%%%%%%%%%%%%%%%%%%%%%%%%%

\end{document}